\documentclass[final,5p,times,twocolumn,authoryear]{elsarticle}

\usepackage{amssymb}
\usepackage{amsmath}
\usepackage{lineno}
\usepackage{siunitx}
\usepackage{textcomp}
\usepackage[dvipsnames]{xcolor}
\usepackage{hyperref}
\hypersetup{breaklinks=true}

\journal{Icarus}

\begin{document}

\begin{frontmatter}



\title{Evolution of granular salty ice analogs for Europa: Sublimation and Irradiation} %% Article title


\author[wp]{Rafael Ottersberg} %% Author name
\author[wp]{Antoine Pommerol} %% Author name
\author[wp]{Linus Leo Stöckli} %% Author name
\author[wp]{Lorenzo Obersnel} %% Author name
\author[wp]{André Galli} %% Author name
\author[iap]{Axel Murk}
\author[wp]{Peter Wurz}
\author[wp]{Nicolas Thomas} %% Author name

\affiliation[wp]{organization={Space Research and Planetary Sciences, Physics Institute, University of Bern},%Department and Organization
            addressline={Siedlerstrasse 5}, 
            city={Bern},
            postcode={3012}, 
            state={Bern},
            country={Switzerland}}
            
\affiliation[iap]{organization={Institute of Applied Physics, University of Bern},%Department and Organization
            addressline={Sidlerstrasse 5}, 
            city={Bern},
            postcode={3012}, 
            state={Bern},
            country={Switzerland}}

%% Abstract
\begin{abstract}
We study the evolution of the Vis-NIR reflectance spectrum of salty granular ice analog samples in a simulation chamber under conditions relevant to the surface of Europa. A novel application and custom calibration of a thermopile sensor enabled the measurement of the surface temperature of the samples in far infrared emission. This allows the kinetics of the observed changes to be scaled to equivalent timescales on Europa. We observed significant changes in the depth and shape of the broad water absorption bands for all samples on timescales of a few thousand years of equatorial conditions on Europa. This effect should be taken into account if quantitative predictions about bulk composition are made based on remote-sensing data. A narrow absorption feature attributed to hydrohalite formed during the sublimation of the sodium chloride sample. We used near-infrared spectroscopy in an irradiation chamber to assess the stability of this narrow feature under electron irradiation. We find that the radiation environment present on Europa dehydrates the hydrohalite on short timescales. Therefore, we expect hydrohalite not to be present on the surface, unless erupted very recently ($<10$ yr) or located in thermal anomalies ($>$ 145 K). Thus, a detection of hydrohalite would clearly indicate recent activity. 
\end{abstract}

%Graphical abstract

%% Keywords

\end{frontmatter}




\section{Introduction}
\label{intro}
Europa, the second of the four Galilean Moons, is a differentiated body with a metal and silicate bearing core and a global ocean of liquid water under an ice shell \citep{anderson_europas_1998, gomez_casajus_updated_2021}. A subsurface ocean and current geological activity make Europa potentially habitable \citep{chyba_astrobiology_2005}. No lander mission has flown to the icy satellites, making remote sensing in the visible and near-infrared wavelength range one of the most valuable tools to constrain the composition of such icy surfaces and thus assess the potential of harboring life underneath. In the following decade, both the MISE onboard Europa Clipper and the MAJIS onboard JUICE will map Europa's surface in the Vis-NIR spectral range with unprecedented spatial resolution \citep{blaney_mapping_2024, poulet_moons_2024}.

Various processes shape the icy surface. Europa is located in Jupiter's magnetosphere and at the outer edge of Io's plasma torus, leading to a harsh radiation and charged particle environment bombarding the surface. \citet{cooper_energetic_2001} reported a UV-C flux (100 - \SI{280}{\nano\meter}) of \SI{4e10}{\keV \centi\meter^{-2} \second^{-1}} and, based on data from EPD onboard Galileo, a charged particles flux of \SI{8e10}{\keV \centi\meter^{-2} \second^{-1}} of which \SI{6e10}{\keV \centi\meter^{-2} \second^{-1}} is due to energetic electrons. \citet{szalay_oxygen_2024} reported the omni-directional electron intensity measured by JADE-E onboard JUNO during the flyby of Europa from \SI{30}{\electronvolt} to \SI{40}{\kilo\electronvolt}, and modeled the extended spectrum down to \SI{1}{\electronvolt}. The integrated flux of the spectrum with the highest intensity is \SI{3e10}{\keV \centi\meter^{-2} \second^{-1}}. This radiation leads to the sputtering of H$_2$O molecules from the ice, a process estimated to ablate approximately \SI{0.2}{\micro\meter} yr$^{-1}$. In comparison, meteorite and micrometeorite impacts are estimated to process the top \SI{2.6}{\meter} of regolith over the last 50 million years, making sputtering the more important process for all but the very youngest surfaces \citep{cooper_energetic_2001, szalay_oxygen_2024}. \citet{nordheim_preservation_2018} showed that the radiation dose rates are highly dependent on surface location, with the highest rates within 'radiation lenses' centered on the leading and trailing hemisphere.
\citet{spencer_thermal_1987} proposed a process called thermal segregation caused by the sublimation of water from the surface and the resulting increase in abundance of darker residual material on the surface.
Because of the lower albedo of this darker material, it absorbs more solar radiation and warms up, causing positive feedback.

Observations of the Near Infrared Mapping Spectrometer (NIMS) on the Galileo spacecraft provided important insights into the surface composition. Besides the water ice, sulfuric acid hydrate has been proposed as a component of the surface \citep{carlson_sulfuric_1999, carlson_distribution_2005}. Furthermore, hydrated salts, especially hydrated magnesium sulfate, have been proposed to be present on the reddish-brown terrains. However, the spectral and spatial resolution of NIMS was insufficient to identify or distinguish between these hydrated salts \citep{dalton_spectral_2003}. \citet{brown_salts_2013} obtained spectra of Europa's leading and trailing hemisphere with an adaptive optics system at the Keck Observatory, with a spectral resolution 40 times better than the NIMS data. For the trailing hemisphere, they present a good spectral match of sulfuric acid hydrate below \SI{1.8}{\micro\meter} and a newly detected \SI{2.07}{\micro\meter} feature that they attributed to hydrated MgSO$_4$. For the leading hemisphere, they reported distorted water ice bands and ruled out many sulfate salts. In a more detailed analysis \citep{fischer_spatially_2015}, three compositionally distinct regions were identified: the bullseye on the trailing hemisphere, the chaos terrains on the leading hemisphere, and the spectrally icy high latitudes. NaCl was reported as one possible alternative to MgSO$_4$ since it is spectrally flat in its anhydrous form. \citet{ligier_vltsinfoni_2016} modeled observations of the SINFONI spectrometer on the VLT with a linear mixing approach and a large spectral library of experimental data. They reported sulfuric acid and Mg-bearing chlorinated salts on the trailing hemisphere. Their modeling showed water ice to be most abundant with grain sizes of 25 - \SI{200}{\micro\meter}.
\citet{trumbo_sodium_2019} observed an absorption feature at \SI{450}{\nano\meter} and \citet{trumbo_new_2022} observed a feature in the mid-UV using the Hubble space telescope that is indicative of irradiated sodium chloride (NaCl).  \citet{tan_spatially_2022} observed Europa with the Subaru/IRCS telescope from \SIrange{1}{1.8}{\micro\meter}, and based on the absorption features around \SI{1.2}{\micro\meter}, they calculated upper limits in the range of 10\% abundance for various Cl-bearing hydrated salts, excluding NaCl.

To enable the interpretation of these observations, different experimental campaigns studied the optical properties of relevant analogs in the laboratory. \citet{grundy_near-infrared_1999} and \citet{mastrapa_optical_2008} measured the optical constants of crystalline and amorphous water ice for varying cryogenic temperatures. \citet{carlson_sulfuric_1999} measured the reflectance spectra of different hydration states of sulfuric acid. \citet{dalton_low_2012} measured the optical constants of hydrated magnesium sulfate. \citet{hanley_reflectance_2014} reported the reflectance function of different hydration states of various chlorine salt species, such as sodium and magnesium chloride. 
\citet{hibbitts_color_2019} measured the formation of color centers of various salts at ambient temperature in the visible spectral range. \citet{denman_influence_2022} measured the visible reflectance spectrum of irradiated sodium chloride at cryogenic temperatures and found that photobleaching leads to a buildup of the strength of the F-centers during the night and a decay caused by insolation during the day on Europa. \citet{brown_mid-uv_2022} measured the mid-UV reflectance spectrum of irradiated NaCl also at temperatures expected on the surface of Europa and observed the formation of a \SI{220}{\nano\meter} absorption feature. \citet{cerubini_vis_2022} measured the visible reflectance spectrum of different electron-irradiated mixtures of water-ice and sodium chloride and reported the formation of F-centers, M-centers, and Na-colloid depending on the sample microstructure and irradiation energy. \citet{vu_probing_2020} used Raman and X-ray diffraction to investigate which salt minerals precipitated when Na$^+$, Mg$^{2+}$, Cl$^-$ and SO$_4^{2-}$ bearing brines freeze and found that in addition to NaSO$_4$ and MgCl$_2$ hydrates, predicted by the chemical divide model, NaCl and MgSO$_4$ frequently form.  In the flash-frozen samples MgSO$_4$ hydrates were found in a vitreous form, which can hinder their detection. \citet{foxpowell_partitioning_2021} used optical and Cryo-SEM microscopy techniques to investigate the partitioning of salt bearing fluids during freezing. They found that under flash-freezing conditions the ice and brine phases segregate with the latter forming a vein network with dimensions below \SI{1}{\micro\meter} within the particles.

Commonly, spectral modeling of observations starts from a spectral library with such endmembers and searches for the best fit to the data \citep{fischer_spatially_2015, ligier_vltsinfoni_2016, king_compositional_2022}. 
In this work, we take a different experimental approach. We begin with water solutions with different salt species dissolved, namely MgSO$_4$, MgCl$_2$, and NaCl. We then produce analog ice samples by flash-freezing droplets and we let them evolve in conditions relevant to the surface of Europa and observe changes in the reflectance spectra (\SIrange{400}{2500}{\nano\meter}). \citet{cerubini_near-infrared_2022} measured the changes of the reflectance spectra of salty particulate ices and ice slabs in a vacuum and reported the change of spectral criteria during sublimation. Building upon this work, the current study adds the capability to measure the surface temperature of the sample, allowing us to assess the kinetics of the sublimation. A particular focus of this work lies in the formation and stability of hydrated forms of NaCl. In Section \ref{sec:methods}, we present the experimental techniques, focusing on a novel method to measure ice surface temperature in far infrared (FIR) emission. In Section \ref{sec:results} results of the sublimation experiments with NaCl, MgSO$_4$, and MgCl$_2$ ice analogs and the irradiation of NaCl ice analog are presented. Section \ref{sec:discussion} discusses the results of the experiments and their implication for remote-sensing observations. Section \ref{sec:summary} gives a short summary of the work.

\section{Material and Methods}
\label{sec:methods}

\subsection{Analogs}
\label{subsec: analogs}
\begin{figure}
    \centering
    \includegraphics[width=\linewidth]{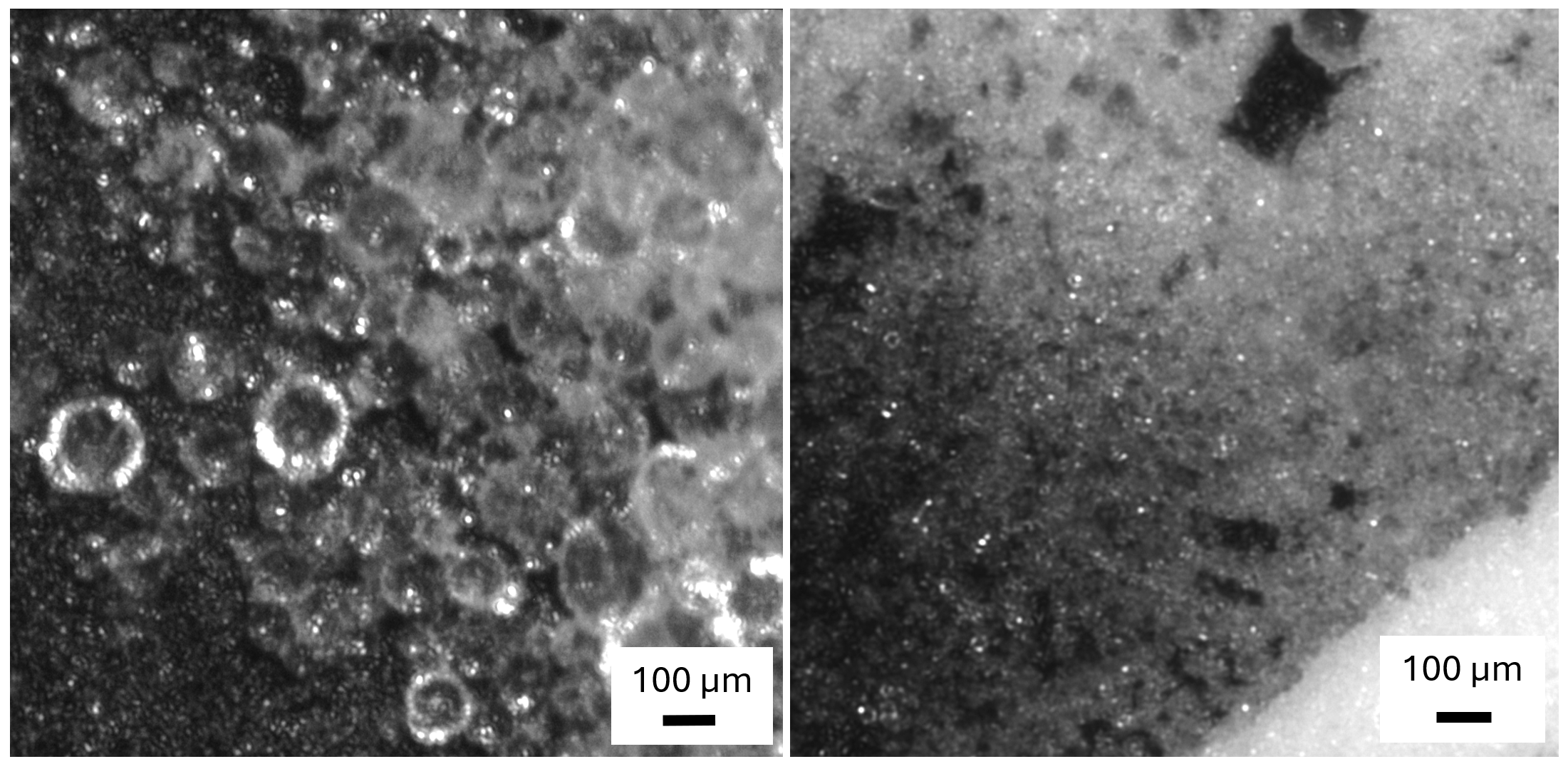}
    \caption{Microscopy images of the granular ice analogs. They were imaged with a long-distance microscope at \SI{77}{\kelvin} and under a nitrogen atmosphere. Left: SPIPA-B (ultrasonic disperser), Right: SPIPA-C (air pressure atomizer)}.
    \label{fig:LDM_SPIPA}
\end{figure}

In this study, two different production protocols for granular ice samples \citep{pommerol_experimenting_2019} have been used. The protocols start with deionized water with the possibility of adding soluble salts to the solution. The first protocol, SPIPA-B, uses an ultrasonic disperser (Hielscher UP200st) that produces spherical water droplets with a mean diameter of 70$\pm$\SI{30}{\micro\meter}. SPIPA-C uses an air pressure atomizer working at \SI{1000}{\hecto\pascal} and produces droplets with a wider size distribution (2-\SI{100}{\micro\meter}). In both protocols the droplets fall into a liquid nitrogen container where they flash-freeze with a cooling rate of tens of Kelvin per second. After sample production, when nearly all LN2 is evaporated, the granulate ice is transferred to the precooled sample holder and the surface smoothed with a precooled tool. The thickness of the sample in the sample holder was measured with a precooled ruler. The sample holder is then mounted in the vacuum chamber, which is purged with nitrogen to avoid any frost deposition on the sample. During this insertion process, the temperature of the ice sample does not exceed  \SI{120}{\kelvin}. Microscopy images taken with a long-distance microscope are shown in Figure \ref{fig:LDM_SPIPA}. The porosity of the SPIPA-B sample is approximately 50\% \citep{pommerol_experimenting_2019}.
\subsection{Sublimation chamber}
\label{subsec:sciteas2}
\begin{figure}
    \centering
    \includegraphics[width=\linewidth]{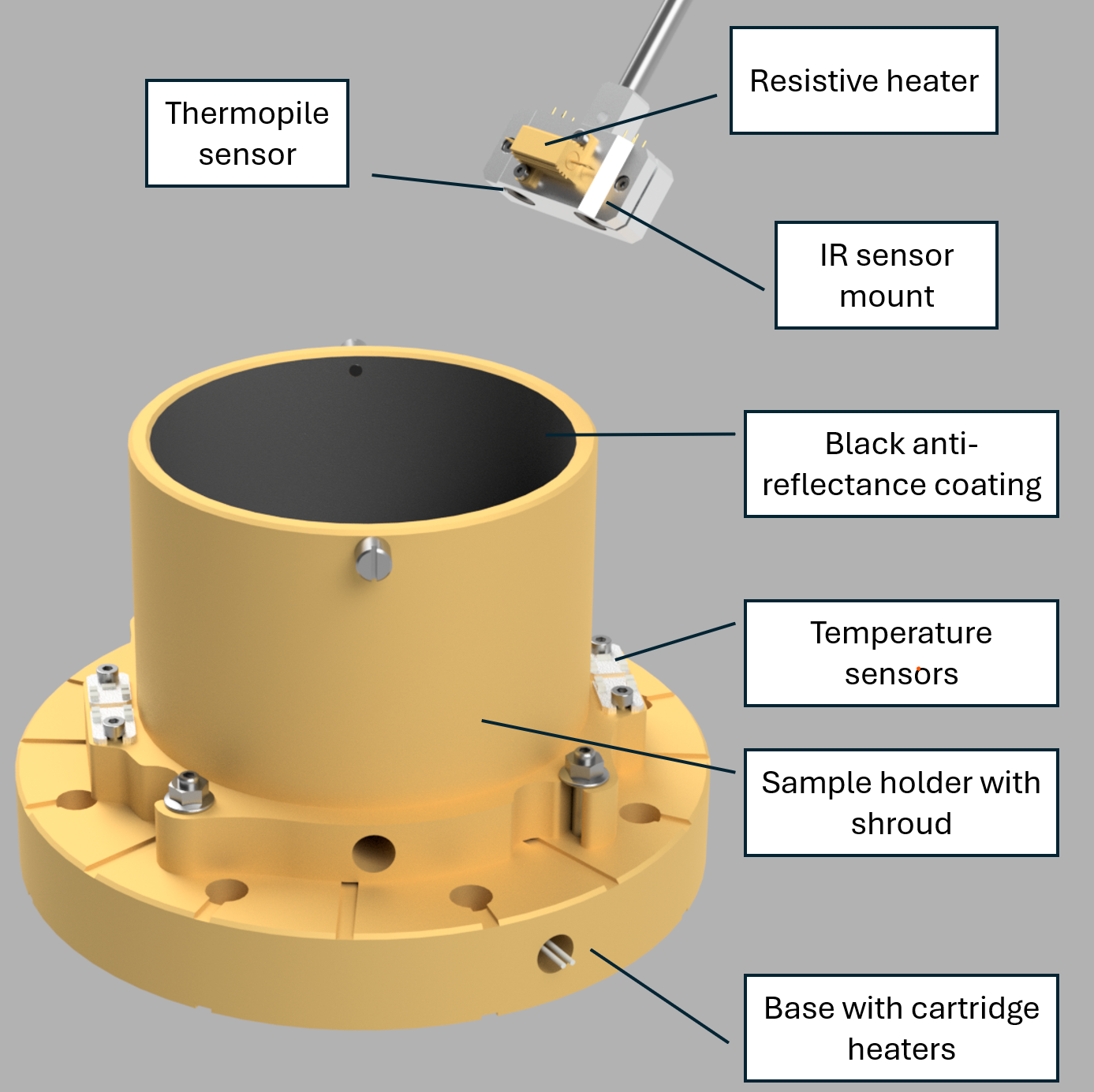}
    \caption{Schematic of the newly developed base and sample holder assembly. The base is bolted to the cryocooler below. Two cartridge heaters are used to control the temperature. The sample holder is removable and is attached with a bayonet connection to the base. The sample holder has multiple temperature sensors bolted down to measure the temperature of the copper precisely. There is a shroud with a diameter and height of \SI{10}{\centi\meter} with a black coating on the inside to prevent reflections of the light source on the sample. The FIR-sensor is mounted on the side looking onto the sample with an angle of approximately 40\textdegree.}
    \label{fig:sample-holder-assy}
\end{figure}
The experiments were conducted in an upgraded version of the Simulation Chamber for Imaging the Temporal Evolution of Analogue Samples 2 (SCITEAS-2) \citep{cerubini_near-infrared_2022}. The chamber has a pumping system consisting of a primary multi-stage roots pump (Pfeiffer ACP15) and a secondary turbomolecular pump (Pfeiffer HiPace300), reaching pressures $<$\SI{1e-4}{\pascal}. %%%The cooling system consists of two independent systems:
%%%\begin{itemize}
    A helium cryo-cooler is attached to a copper base plate on which the sample holder is bolted down. This ensures good thermal conductivity between the sample holder and the baseplate, despite vibrations from the cryo-head. A new sample holder and base plate assembly, shown in Figure \ref{fig:sample-holder-assy}, was designed and manufactured. Two \SI{200}{\watt} Cartridge heaters are placed in the gold-coated copper support plate to compensate for the constant cooling of the cryocooler. A newly developed electronics unit reads out the temperature sensors and controls the heating by running a proportional-differential-integral (PID) feedback loop. The sensors are Pt-100 and Pt-1000 resistance temperature detectors (RTDs) and diodes (BAS16, \citet{rijpma_cryogenic_2006}), which are mounted on an aluminum circuit board  \citep{stockli_metal_2024} and bolted to the sample holder. While the RTDs perform well above \SI{50}{\kelvin} with a general calibration, the diodes have better sensitivity at lower temperatures with a custom calibration.  The gradient between the top rim of the sample holder and the bottom where the sensors are attached, is less than \SI{0.2}{\kelvin} due to the good thermal conductivity of the copper. The temperature of the sample holder can be controlled precisely ($\pm$\SI{0.5}{K}) in the wide temperature range of 30 to \SI{450}{\kelvin}. 

    %%%\item Optionally, a large liquid nitrogen shroud around the sample holder can be used to reduce the exposure of the sample to infrared radiation from the warm environment. A solenoid valve was added to control the cooling of the shroud, and a copper coil tube in a water bath was added on the exhaust side to preheat the gas before it entered a ventilation shaft, resulting in an automated, regulated and safe process.
%%%\end{itemize}

The circular sample holder has a wall with a height of \SI{10}{\centi\meter} and a diameter of \SI{10}{\centi\meter}. The high wall shields the sample from a large part of the ambient thermal radiation. 
A quartz window with a diameter of \SI{13}{\centi\meter} located \SI{40}{\centi\meter} over the sample allows the measurement of the optical properties of the analog samples in reflectance in the visible (Vis) and near-infrared (NIR) range. The hyperspectral imaging system is described in Section \ref{subsec:hyperspecimaging}. A polished golden lid can be slid over the top of the sample to shield it from all ambient radiation emitted from the warm chamber walls and window. In this study, this functionality was used to estimate the amount of reflected thermal radiation on the sample. During the sublimation experiment the lid remained open.
Additional RTDs can be read out using a Keithley 2700 Multimeter with a multiplexer. All measurements are taken with a 4-wire measurement technique to remove any effects of the resistance of the cables. A PT100 with thin wires is used to measure the temperature inside of the sample. The PT100 is chosen because of lower self-heating compared to a PT1000. Extra thin copper wires (\o = \SI{0.16}{\milli\meter}) are used and a \SI{10}{\centi\meter} long loop is laid in the sample to minimize heat influx from the walls of the sample holder to the sensor. A small piece of black aluminum tape is attached to the sensor to maximize the thermal contact with the porous ice. A thermopile sensor in a thermally stabilized mount (see Section \ref{subsec:fir-measurement}) is placed above the sample to measure the thermal infrared emission.

To measure automatically and reliably for weeks, all devices are connected to a single-board computer (Raspberry Pi 4), which constantly logs all metadata to a database server (InfluxDB) and controls the pumps, helium-cryocooler, liquid nitrogen cooling, gauges, and valves. The control program and all device drivers to communicate with the experiment hardware are written in Python and are accessible as open-source software \citep{sciteas-2-ctrl}.

\subsection{Irradiation chamber}
\label{subsec:mefisto}
The irradiation experiments were conducted in the MEFISTO (MEsskammer für FlugzeitInStrumente und Time-Of-Flight) facility at the University of Bern.  The facility consists of a large vacuum chamber (2 m$^{3}$) including a liquid nitrogen-based cooling stage for the ice samples and a mass spectrometer \citep{galli_surface_2016, tinner_electroninduced_2024}. For the current study, we performed irradiation with the electron gun at 2~keV energy (electron gun, manufacturer: Kimball Physics).
%%% MEFISTO offers the option to irradiate samples with ions from an electron cyclotron resonance ion source \citep{marti_calibration_2001}, electrons (electron gun, manufacturer: Kimball Physics), and UV light (broadband UV lamp or Ly-$\alpha$). 

\subsection{Surface temperature measurement}
\label{subsec:fir-measurement}
Achieving precise temperature control of porous samples in vacuum is challenging due to the poor thermal conductivity caused by the lack of convection. Even if the temperature of the sample holder can be controlled precisely, there can be a large thermal gradient over the sample. Only the uppermost layers (\si{\micro\meter}-\si{\milli\meter}) are probed in the Vis and NIR range. Therefore, a technique to measure the surface temperature of the ice is needed. We mounted and calibrated a commercial thermopile sensor (Melexis MLX90614, \SI{5}{\degree}  field of view) inside the vacuum chamber. Such thermopile sensors are frequently used in commercial products, such as non-contact thermometers, as well as space applications on both orbiters and landers like the MUPUS instrument \citep{spohn_mupus_2007} on Rosetta or the REMS instrument \citep{gomez-elvira_rems_2012} on the Mars Curiosity Rover. Such sensors are readily used down to temperatures of \SI{237}{\kelvin}. To achieve accurate surface temperature measurements down to cryogenic temperatures, we developed a custom sensor mount with an integrated heating system shown in Figure \ref{fig:sample-holder-assy} and a calibration algorithm. To stabilize the thermopile sensor temperature at \SI{300}{\kelvin} $\pm$ \SI{0.05}{\kelvin}, the onboard temperature sensor of the Melexis package is read out, and a PID-feedback loop controls a resistive heater. \ref{sec:thermopile-calibration} describes the developed calibration procedure and calibration algorithm in detail.

The biggest source of uncertainty of the surface temperature measurement comes from the assumption that the emissivity of the sample is equal to the emissivity of the calibration standard.  Especially, because the emissivity of the sample can change during the sublimation experiment. Both the microstructure and the composition of a surface impact its emissivity. A decreasing emissivity has two effects when working at cryogenic temperatures in a vacuum chamber. Firstly, the intensity of FIR radiation emitted from the sample is proportional to the emissivity $\epsilon$ and decreases. Secondly, the intensity of ambient background radiation from windows and chamber walls reflected on the sample is proportional to $1-\epsilon$ and increases (see Section \ref{sec:thermopile-calibration}). In the temperature range we studied in the sublimation experiment (\SIrange{160}{185}{\kelvin}) these two effects compensate each other. The measurement error for a sample with an emissivity $\epsilon=0.8$ when calibrated against a standard with $\epsilon=0.98$ is $<$\SI{3}{\kelvin} at $T=$\SI{165}{\kelvin} and $<$\SI{0.5}{\kelvin} at $T=$\SI{180}{\kelvin}. 

\subsection{Hyperspectral imaging system}
\label{subsec:hyperspecimaging}
The optical measurements were performed with an upgraded version of the Mobile Hyperspectral Imaging System (MoHIS) \citep{cerubini_near-infrared_2022}. The system comprises a broadband light source (Newport QTH), a newly installed monochromator (QD MSH300), a custom-made large fiber bundle (CeramOptec\textregistered), a Vis camera (Thorlabs Scientific CCD), and a NIR camera (Xenics Xeva 320; MCT sensor). A new near-infrared lens (Stingray Cherry 50mm f/1.3) was installed with a narrower field of view. The two cameras see the sample through two mirrors that are 45\textdegree\ tilted. The phase angle is approximately \SI{3}{\degree} in the sublimation chamber and \SI{2}{\degree} in the irradiation chamber.
 The monochromator has two variable slits. The slit widths were set to \SI{1}{\milli\meter}, which corresponds to a bandwidth of \SI{5.4}{\nano\meter} below and \SI{10.8}{\nano\meter} above \SI{2}{\micro\meter}.
The measurement program was rewritten to speed up the acquisition and allow the automated control of experimental parameters of the simulation chamber. A hyperspectral image cube (\SIrange{400}{2500}{\nano\meter}, \SI{5}{\nano\meter} steps) can be acquired in 15 min.

The NIR camera takes all images with a constant exposure time. It takes 8 dark exposures (light source shutter closed) spread through the cube acquisition. These dark exposures are subsequently interpolated and subtracted from the sample images so that small changes of the dark current caused by drifts of the sensor temperature are calibrated. The Vis uses different exposure times in the range of \SI{200}{\milli\second} to \SI{5}{\second}. Dark images with different exposure times are taken at the beginning and at the end of the visible cube. The two image sets are interpolated, also in exposure time, and subsequently subtracted from the sample image. This dark subtraction is performed for the science images and the flatfield images. The flatfield standard is a Spectralon\textregistered\ target with nearly Lambertian scattering properties. A shallow absorption feature of the Labsphere Spectralon\textregistered\ around \SI{2}{\micro\meter} is corrected using a reflectance spectrum provided by the manufacturer. The science images for every wavelength are then divided by the flatfield images at the corresponding wavelength. 

Hot or cold pixels on the MCT image sensors are excluded with a bad pixel mask. The bad pixel mask was generated by imaging a flatfield target that filled the whole field of view of the sensor. Exposures with two different illuminations were taken and the dark signal was removed. Then, the ratio of these two images with different exposure times was calculated. All 5 sigma outliers in this ratio-image were identified as bad pixels. This amounted to 352 pixels on the 256x320 pixel sensor.

Finally, regions of interest (ROI) are defined by drawing polygons on the sample's image and averaging all pixels within this shape. For every defined ROI, only the 2 to 98\% percentiles are used to calculate the mean reflectance to remove any outlier pixels.

Alternatively, spectral ratios from the image cubes can be calculated, allowing the detection of spatially localized spectral changes.

\subsection{Scaling of sublimation kinetics}
\label{subsec:scaling}
We scaled the sublimation kinetics observed in the simulation chamber to conditions on the surface of Europa at low latitudes. 
For the scaling, we used a form of the Hertz-Knudsen equation from \citet{kossacki_temperature_2014}:
\begin{equation}
    \frac{\mathrm{d}z}{\mathrm{d}t}(T) = \frac{B\alpha}{\rho} \sqrt{\frac{\mu}{2 \pi R_g T}}(p_{sat}-p), 
\end{equation}
where $B$ and $\alpha$ are the dimensionless return flux and sublimation coefficients,  $\rho$  the density of the subliming material, $\mu$ the molar mass of ice, $R_g$ the universal gas constant, $T$ the absolute temperature of the material, $p$ the water vapor pressure over the subliming surface and $p_{sat}$ the saturation pressure.
We use the saturation pressure function, given in \citet{wagner_new_2011}, calculated as
 \begin{equation}
     p_{sat}(T) = p_t\times\exp\left(\theta^{-1} \sum_{i=1}^{3}a_i \theta^{b_i}\right)
     ,
 \end{equation}
 where $\theta = T / T_t$. $T_t = \SI{273.16}{\kelvin}$ and $p_t=\SI{611.657}{\pascal}$ are the temperature and pressure at the triple point of water. The coefficients $a_i$ and $b_i$ are given in Table \ref{tab:coeff}. We chose to use this saturation pressure function following the argument in \citet{macias_molecular_2023}.

\begin{table}
\begin{tabular}{lrl}
\hline
i & \multicolumn{1}{c}{$a_i$}    & \multicolumn{1}{c}{$b_i$}      \\ \hline
1 & $-0.212 144 006\times10^2$ & $0.333 333 333\times10^{-2}$ \\
2 & $0.273 203 819\times10^2$   & $0.120 666 667\times10^1$  \\
3 & $-0.610 598 130\times10^1$  & $0.170 333 333\times10^1$  \\ \hline
\end{tabular}
\caption{Coefficients for the saturation pressure function reproduced from \citet{wagner_new_2011}.}
\label{tab:coeff}
\end{table}
 
The total recession during the experiment is numerically integrated using the pressure and surface temperature continuously measured  ($\Delta t = \SI{5}{\second}$) in the simulation chamber. Subsequently, the result is divided by the average recession rate calculated for the surface at the equator of Europa.
To this end, a model for the sublimation based on a measured diurnal brightness temperature profile at the equator of Europa \citep{pappalardo_surface_2009} was created. An emissivity of 0.95 to calculate the surface temperature from the brightness temperature is assumed.  The diurnal surface temperature profile is lowest at \SI{87}{\kelvin} during the night and peaks at \SI{132}{\kelvin} during the day. The vapor pressure $p$ at the surface is assumed to be negligible, when compared to the saturation pressure $p_{sat}$. The integrated sublimation per European day amounts to \SI{6.7e-4}{\micro\meter} and the averaged recession rate is \SI{0.07}{\micro\meter}/yr. We note that over 90\% of the total sublimation happens when the sun is within $\pm30$\textdegree of the zenith. Using this average sublimation rate, we can calculate how many years of equatorial conditions on Europa are needed to observe a similar change as measured in the sublimation chamber. In the division of the recession rates, the parameters $B$, $\alpha$ and $\rho$ cancel out.  An example of such a scaling is shown in Figure \ref{fig:pressure-temp-time}. For the sublimation values provided above $B=0.75$, $\alpha=1$ and $\rho=\SI{0.92}{\gram \cm^{-3}}$ were used. The error of \SI{3}{\kelvin} of the surface temperature measurement in the vacuum chamber translates to a factor of 2 in the calculated sublimation time, showing that the calculated sublimation timescales are a rough estimate.
Since the observed surface brightness temperature of Europa strongly decreases with latitude, the calculated sublimation timescales should be interpreted as a lower limit. Conversely, local thermal anomalies, such as hotspots would accelerate the sublimation process significantly. An increase of $\Delta T=\SI{15}{\kelvin}$ accelerates the sublimation by 2 orders of magnitude.
 
\section{Results}
\label{sec:results}
\subsection{Sublimation experiments}
\begin{figure}
    \centering
    \includegraphics[width=\linewidth]{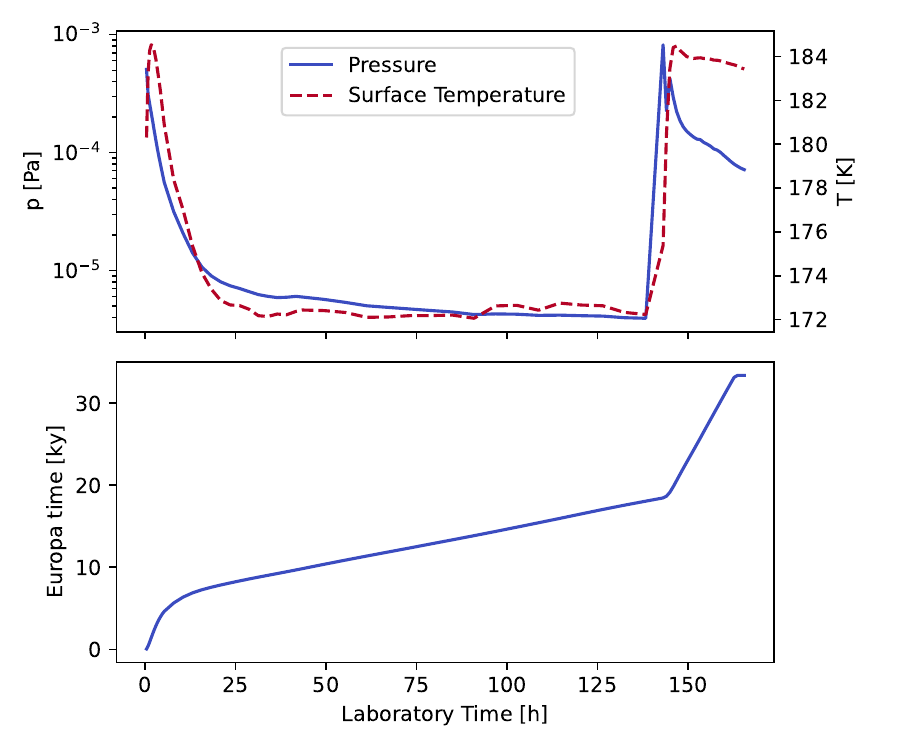}
    \caption{The top panel shows the measured simulation chamber pressure and surface temperature of the MgCl$_2$ sample during the sublimation experiment. The bottom panel shows the scaling from the Laboratory Time in hours to the Europa Time in 1000 years (ky). The scaling is based on the measured pressure and surface temperature and assumes a diurnal temperature profile measured at the equator of Europa ($T_{peak}$=132K).}
    \label{fig:pressure-temp-time}
\end{figure}

The sublimation experiments were conducted in two phases. In a first phase the sample holder was set to a temperature of 45-50 K, and the samples sublimated with a surface temperature of 160 - 180 K depending on the sample thickness and the thermal conductivity of the sample. In a second phase the sample holder was heated up to 120 - 130 K and the samples sublimated at an increased rate with a surface temperature of 175 - 185 K. In the first phase, the pressure in the chamber was approximately \SI{5e-6}{\pascal}. In the second phase, the pressure was in the range of \SIrange{1e-4}{1e-3}{\pascal}. As an example, the pressure and surface temperature measurement of the MgCl$_{2}$ experiment, together with the scaled time as described in Section \ref{subsec:scaling} are shown in Figure \ref{fig:pressure-temp-time}. A temperature sensor as described in Section \ref{subsec:sciteas2} was placed in all samples. The depth of this sensor was measured after the experiment, when the sample was extracted from the chamber and the sensor was dug out of the ice. It is only a rough estimate and it cannot be ruled out that the sensor moved in this process.

\subsubsection{Sodium Chloride}
\label{subsubsec:nacl}
\begin{figure}
    \centering
    \includegraphics[width=\linewidth]{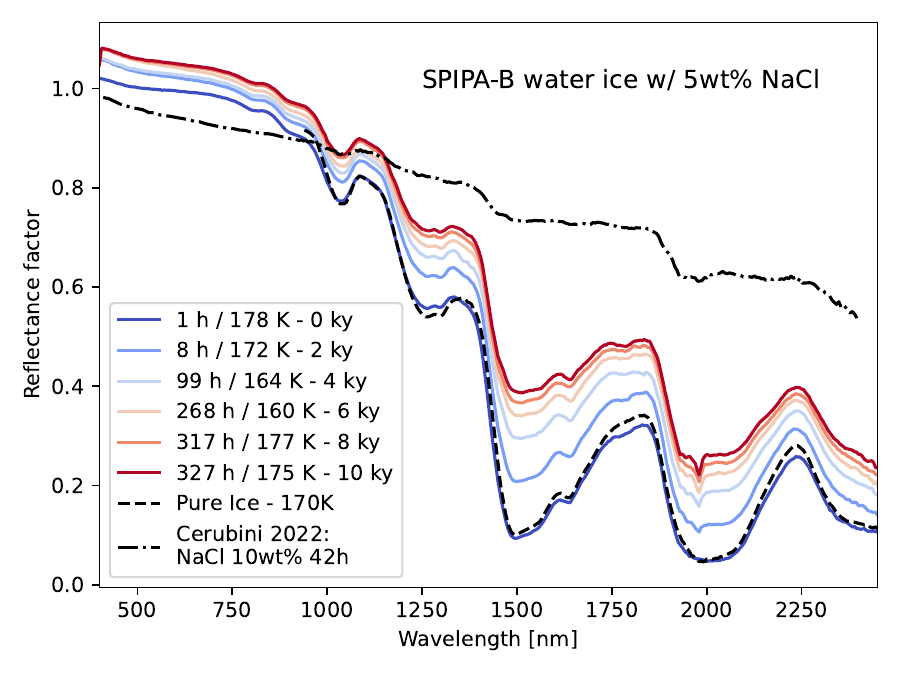}
    \caption{Sublimation of a grainy ice analog with 5wt\% NaCl.  The relative laboratory time and the measured surface temperature for all spectra are shown in the legend.  The estimated timescale in the unit of 1000 years (ky), when the sublimation kinetics are scaled to Europa's equatorial conditions, is given after the dash. The dashed-dotted line shows the final spectrum of the sublimation experiment with a 10wt\% NaCl SPIPA-B analog sample in \citet{cerubini_near-infrared_2022}.}
    \label{fig:nacl-sublimation}
\end{figure}

The circular sample holder was filled with granular ice produced from a 5wt\% NaCl solution using the SPIPA-B protocol. The sample had a thickness of \SI{15}{\milli\meter}, and the surface was flattened with a cold spoon.  The change of the reflectance spectrum during the sublimation experiment can be seen in Figure \ref{fig:nacl-sublimation}. The sublimation time in hours and the measured surface temperatures during the acquisition of the spectral cubes are shown in the legend. This surface temperature is important for the interpretation of temperature-dependent absorption features like the \SI{1.65}{\micro\meter} band. The in-ice temperature sensor was placed 1-\SI{2}{\milli\meter} below the surface of the sample. During the first phase of the sublimation experiment, when the measured surface temperature was between \SI{160}{\kelvin} and \SI{165}{\kelvin}, the sample holder was at \SI{45}{\kelvin}, and the temperature sensor in the ice at \SI{138}{\kelvin}.

At the beginning of the experiment, the reflectance spectrum of the salty ice sample is nearly identical to the pure water ice spectrum. Only a shallow feature around \SI{1.8}{\micro\meter} is present in the NaCl sample and absent in the pure ice sample. As the experiment continues, the reflectance factor over the whole spectral range increases, most prominently in the range of the broad water absorption bands around 1.6 and \SI{2}{\micro\meter}. A narrow absorption feature develops at \SI{1.98}{\micro\meter}, which is not present at the beginning of the experiment. Additionally the shape of the shoulder at \SI{1.8}{\micro\meter} changes, revealing a shallow absorption feature.

\subsubsection{Magnesium sulfate}
\label{subsubsec:mgso4}
\begin{figure}
    \centering
    \includegraphics[width=\linewidth]{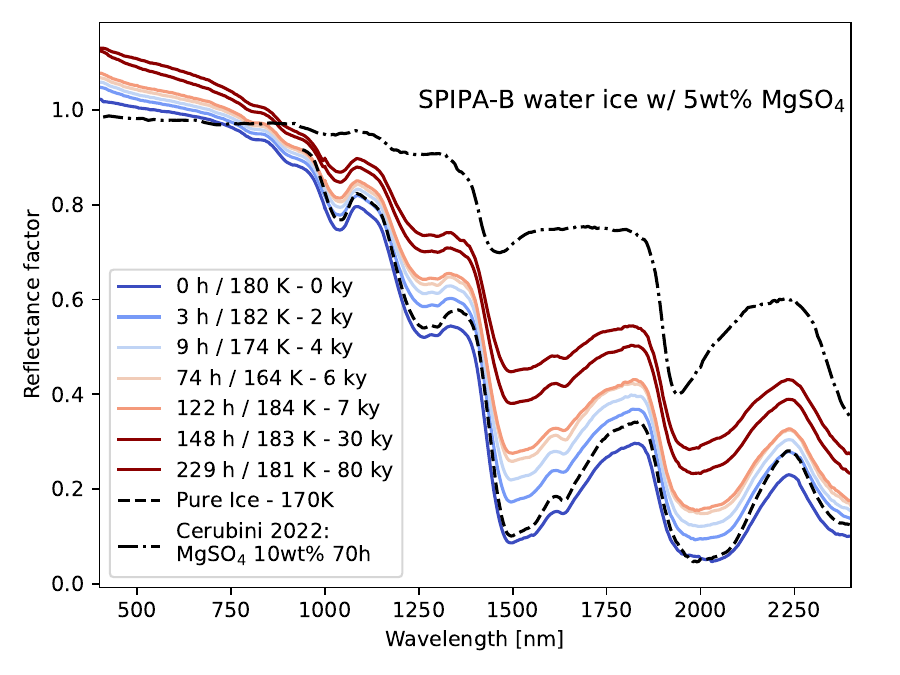}
    \caption{Sublimation of a grainy ice analog with 5wt\% MgSO$_4$.  The relative laboratory time and the measured surface temperature for all spectra are shown in the legend.  The estimated timescale in units of 1000 years (ky), when the sublimation kinetics are scaled to Europa's equatorial conditions, is given after the dash. The dashed-dotted line shows the final spectrum of the sublimation experiment with a 10wt\% MgSO$_4$ SPIPA-B analog sample in \citet{cerubini_near-infrared_2022}.}
    \label{fig:mgso4-sublimation}
\end{figure}

The magnesium sulfate ice sample produced from a 5wt\% MgSO$_4$ solution was approximately 12-\SI{17}{\milli\meter} thick. The in-ice temperature sensor was placed \SI{5}{\milli\meter} below the surface. Figure \ref{fig:mgso4-sublimation} shows that the reflectance in the NIR was slightly lower compared to a pure ice sample. When the sample sublimates, the reflectance increases, most notably in the valleys of the broad water absorption bands. The center of the broad \SI{2}{\micro\meter} band becomes slightly skewed to lower wavelengths. No narrow absorption features are detected during the sublimation. In the first phase of the experiment, the sample holder was at \SI{50}{\kelvin}, the temperature sensor in the ice measured \SI{95}{\kelvin}, and the FIR sensor measured a surface temperature of \SI{165}{\kelvin}.

\subsubsection{Magnesium chloride}
\label{subsubsec:mgcl2}
\begin{figure}
    \centering
    \includegraphics[width=\linewidth]{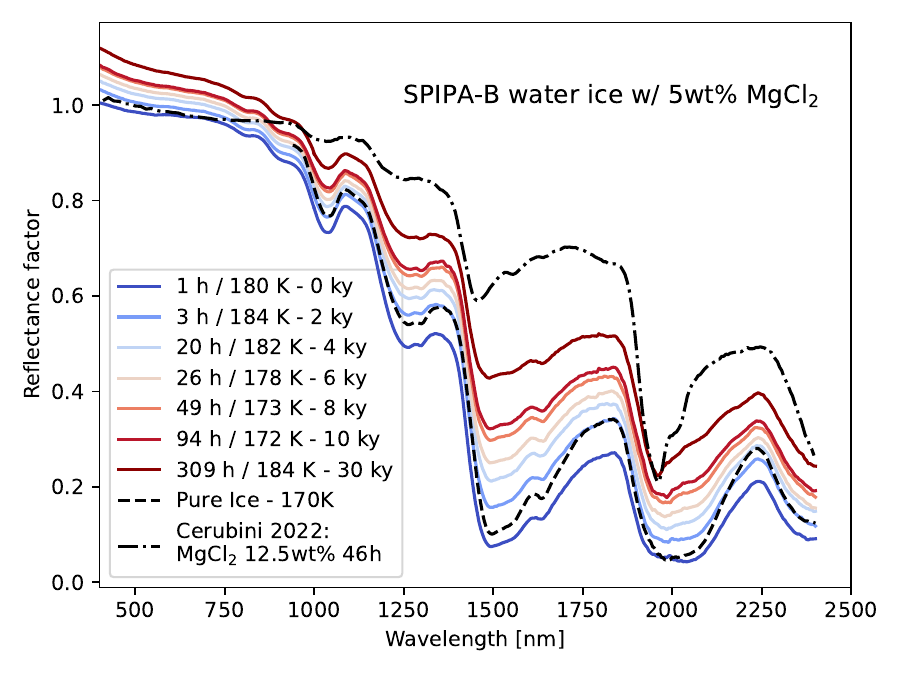}
    \caption{Sublimation of a grainy ice analog with 5wt\% MgCl$_2$. The relative laboratory time and the measured surface temperature for all spectra are shown in the legend.  The estimated timescale in units of 1000 years (ky), when the sublimation kinetics are scaled to Europa's equatorial conditions, is given after the dash. The dashed-dotted line shows the final spectrum of the sublimation experiment with a 10wt\% MgCl$_2$ SPIPA-B analog sample in \citet{cerubini_near-infrared_2022}.}
    \label{fig:mgcl2-sublimation}
\end{figure}

The magnesium chloride ice sample produced from a 5wt\% MgCl$_2$ solution was approximately 15-\SI{18}{\milli\meter} thick. The in-ice temperature sensor was placed \SI{5}{\milli\meter} below the surface. Figure \ref{fig:mgcl2-sublimation} shows that the reflectance in the NIR was lower compared to a pure ice sample. When the sample sublimates, the reflectance in the whole Vis-NIR wavelength range increases, most notably in the valleys of the broad water absorption bands. The \SI{2}{\micro\meter} absorption feature develops a strong asymmetric shape, skewed towards lower wavelength, with a minimum at \SI{1.95}{\micro\meter}. In the Vis, the spectral slope becomes slightly bluer. In the first phase of the experiment, the sample holder was at \SI{52}{\kelvin}, the temperature sensor in the ice measured \SI{160}{\kelvin} and the FIR sensor measured a surface temperature of \SI{173}{\kelvin}.

\subsection{Irradiation experiment}
The formation of the narrow \SI{1.98}{\micro\meter} feature under sublimation, which is not observed in high spectral resolution ground-based observations like the ones presented in \citep{ligier_vltsinfoni_2016} led us to investigate the stability of this feature in a high-radiation environment. A porous ice sample produced from a 5wt\% NaCl solution using the SPIPA-C protocol was put in the MEFISTO chamber. The sample was left for \SI{60}{\hour} at approximately 180-\SI{190}{\kelvin} so the surface sublimated and formed the absorption feature as seen in previous experiments in SCITEAS-2. Subsequently it was irradiated with \SI{2}{\kilo\electronvolt} for \SI{10}{\minute}. The electron current measured on the Faraday cup was \SI{10}{\micro\ampere}. The area of the irradiated spot was \SI{1.1}{\centi\meter^2}. Therefore the total dose on the irradiated ice was \SI{3.4e16}{e^- \centi\meter^{-2}}. Compared to the estimated electron fluxes on Europa's surface of \SI{3e8}{e^- \centi\meter^{-2} \second^{-1}} \citep{cooper_energetic_2001}, this corresponds to a few years of exposure at the surface of Europa. The reflectance spectrum was measured with MoHIS both before and after the irradiation.
\begin{figure}
    \centering
    \includegraphics[width=\linewidth]{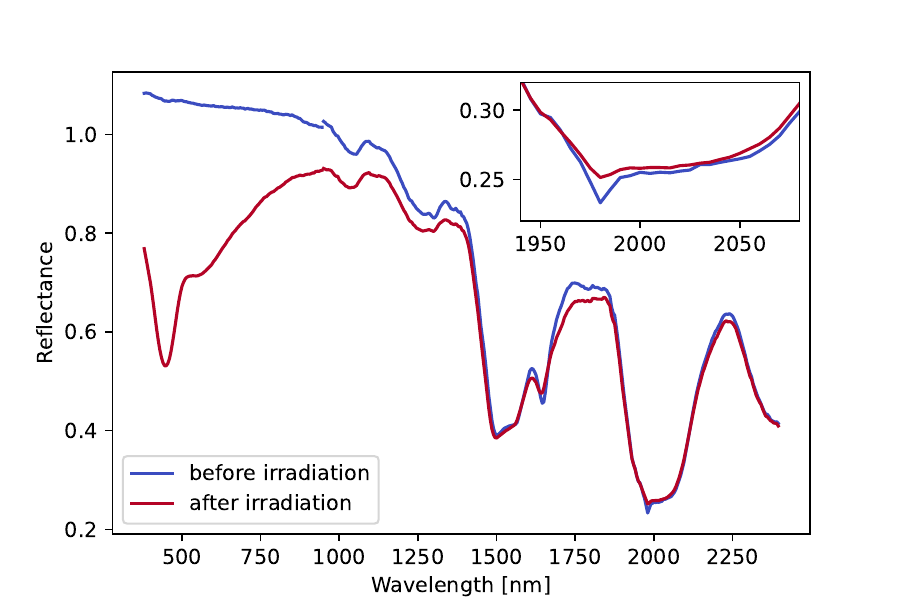}
    \caption{The reflectance spectra of a grainy ice analog (SPIPA-C) with 5wt\% NaCl. The two lines show the reflectance before and after irradiation with \SI{2}{\kilo\electronvolt} for \SI{10}{\minute} with a current of \SI{10}{\micro\ampere}. The irradiation leads to the formation of color centers in the Vis and reduces the depth of the absorption band at \SI{1.98}{\micro\meter}. To improve S/N, 9 spectra before and 9 spectra after the irradiation have been averaged.}
    \label{fig:nacl-irradiation}
\end{figure}

As can be seen in Figure \ref{fig:nacl-irradiation} the sample develops strong absorption features, so-called color centers, in the Vis, namely the F-center at \SI{450}{\nano\meter}, the M-center at $\sim$\SI{720}{\nano\meter} and the broad colloid absorption spanning a wide range up to roughly \SI{1}{\micro\meter}.
The difference in the NIR around the band at \SI{1.64}{\micro\meter} can be explained because the sample was kept at a lower temperature when the ``before measurement`` was done. However, the weakening at \SI{1.98}{\micro\meter} is not related to the temperature change as it is not reversible when the sample is cooled down after the irradiation.

\section{Discussion}
\label{sec:discussion}
\subsection{Thermometry and thermal gradient during sublimation experiment}
The equilibrium temperature of the surface is determined by the energy influx through thermal radiation from the chamber walls and window at ambient temperature on the one hand and the thermal flow through the sample onto the cooled sample holder on the other hand. When the sample holder temperature is set to \SI{45}{\kelvin}, the thermopile sensor measures surface temperatures of \SI{165}{\kelvin}, a difference of more than \SI{100}{ \kelvin} for a sample with a thickness of \SI{15}{\milli\meter}. This demonstrates the poor thermal conductivity of the granular ice sample in a vacuum. This also shows that the sublimation only takes place in the uppermost layer of the sample, the lower part remains at a much lower temperature. While the measurement of the resistive temperature sensors within the sample consistently lay between the measured surface and sample holder temperature, the current measurement setup is limited because the sensor is not fixed in a well-defined position. This is reflected in the large differences of the measured temperatures between the different samples, making quantitative interpretations challenging. This shows the challenge of constraining the surface temperatures with contact temperature sensors on the sample holder or within the ice, stressing the necessity of temperature measurement in emission when working with porous samples in a vacuum.

\subsection{Formation of a salt crust through sublimation}
\begin{figure}
    \centering
    \includegraphics[width=\linewidth]{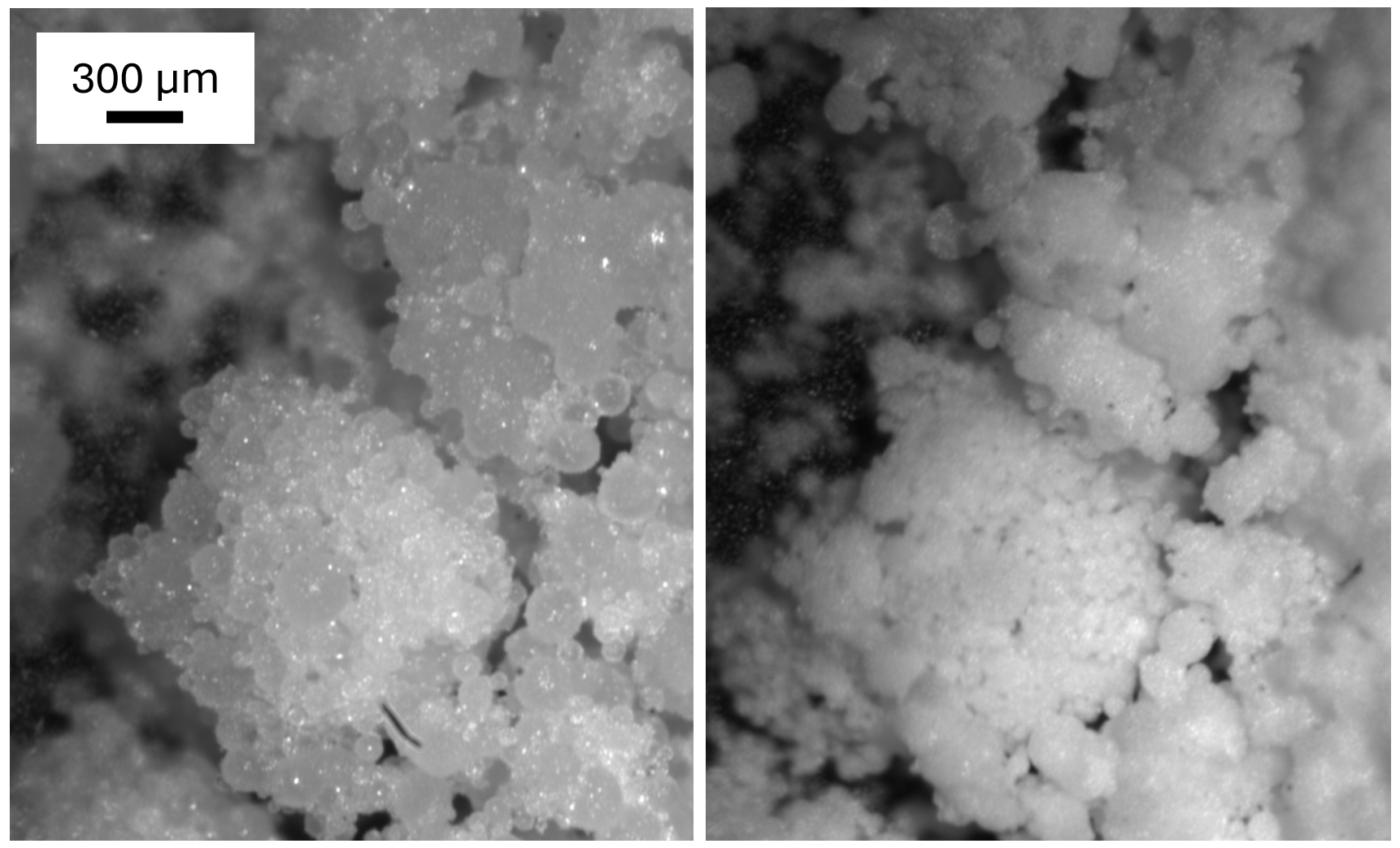}
    \caption{A sample of SPIPA-B ice with 3wt\% NaCl before (left) and after (right) sublimation at high temperatures ($>$\SI{195}{\kelvin}) for 24h imaged in the simulation chamber with a long-distance microscope. Before inserting it into the chamber, the ice was warmed up under the N2 atmosphere, and therefore, the particles formed clumps. These clumps do not disintegrate during sublimation and individual particles stay in place.}
    \label{fig:LDM-NaCl}
\end{figure}

We explain the changes in the optical properties during sublimation with the loss of H$_2$O molecules from the surface of the salty ice particles, through which a salt crust forms, which begins to dominate optically. For illustrative purposes, we let a sample of SPIPA-B - 3wt\% NaCl particles sublimate at high temperatures  $>$\SI{195}{\kelvin} for 24h. Before the sublimation, the sample was warmed up under a nitrogen atmosphere, making the particles stick together, forming large clumps. Note that the samples presented in Section \ref{sec:results} were not warmed up and did not form any clumps. Figure \ref{fig:LDM-NaCl} shows the same part of a sample before and after sublimation.
While the arrangement of the grains stays roughly identical, the optical properties change a lot during the sublimation process. The grains that were partly translucent before the sublimation became fully opaque. If a sample is left in a vacuum at high temperatures, the bulk volume only decreases slightly, leaving a highly porous salt structure nearly or completely free of water. This is consistent with the observation by \citet{foxpowell_partitioning_2021} that flash-frozen liquids with dissolved salts form a vein structure.
Since the albedo of the pure salt species (NaCl, MgSO$_4$, MgCl$_2$) is higher than that of pure water ice, the sublimation increases the albedo of the samples.

\subsection{Evolution of the hydration bands}
\begin{figure}
    \centering
    \includegraphics[width=\linewidth]{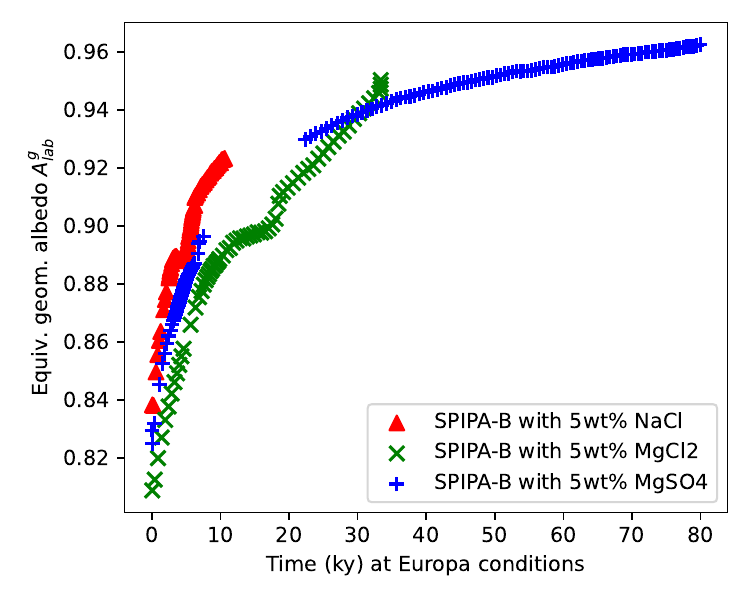}
    \caption{The \textit{equivalent geometric albedo} $A^g_{lab}$ calculated from the measured reflectance spectra during the sublimation experiment. The time in units of 1000 years on the x-axis is scaled to equatorial conditions on Europa using a diurnal brightness temperature profile peaking at \SI{132}{\kelvin}.}
    \label{fig:albedo}
\end{figure}

All samples show a rapid change in the depth of the hydration bands over a few thousand years of thermal day-night cycles on Europa. We calculate a quantity that is called \textit{equivalent geometric albedo} in \citet{beck_low-phase_2021} as
\begin{equation}
    A^g_{lab} = \frac{\int_{\lambda=\SI{400}{\nano\meter}}^{\SI{2500}{\nano\meter}} R_\lambda \times BB_\lambda \,\mathrm{d}\lambda}{\int_{\lambda=\SI{400}{\nano\meter}}^{\SI{2500}{\nano\meter}} BB_\lambda \,\mathrm{d}\lambda}    
\end{equation}
for all spectra of the sublimation experiment. The results are shown in Figure \ref{fig:albedo}. This quantity increases by more than 10\% for all samples sublimating for 10000 yr at Europa's equatorial surface conditions. This shows that the sublimation of salty analogs and the resulting decrease in hydration band depth can significantly impact the radiation balance.

While the shape of the broad water hydration bands in the NaCl analog nearly stays the same over the whole sublimation, the \SI{2}{\micro\meter} band of MgSO$_4$ is skewed slightly, and the one of MgCl$_2$ is skewed strongly towards a shorter wavelength. Especially in the case of MgCl$_2$ this change of the band shape would be detectable in remote sensing data. Figures \ref{fig:nacl-sublimation}-\ref{fig:mgcl2-sublimation} also show the final spectrum of the sublimation experiments conducted at higher temperatures by \citet{cerubini_near-infrared_2022}. The changes in the hydration bands discussed above are consistent with the ones observed in this previous study. For the Mg-bearing samples, the blue slope observed in this study is significantly stronger than in the final data of the previous study. The absence of narrow absorption bands in the MgSO$_4$ sample can be explained by the observation that MgSO$_4$ precipitates in a vitreous form in a flash-frozen solution \citep{vu_probing_2020}.

\subsection{Formation and stability of hydrohalite}
The salty ice analog produced from a 5wt\% NaCl solution has a nearly identical reflectance spectrum as pure H$_2$O ice particles. Only a small absorption feature at the shoulder around \SI{1.8}{\micro\meter} is visible in the pristine sample at the beginning of the experiment. However, as the analog sample sublimates, a narrow absorption feature at \SI{1.98}{\micro\meter} develops, which is not present at the beginning of the experiment. This absorption feature is attributed to hydrohalite (NaCl $\cdot$ 2 H$_2$O). Since the mean optical path length of photons with a wavelength $\lambda=$\SI{2}{\micro\meter} is of the same order as the mean particle diameter of the analog sample, the hydrohalite absorption should be observable if it were present from the beginning of the experiment. Therefore, we argue that hydrohalite forms during the sublimation of the surface. 
This formation of hydrohalite on short timescales on Europa is challenged by the absence of a \SI{1.98}{\micro\meter} feature in all observations to date including the surveys from \citet{brown_salts_2013} and \citet{ligier_vltsinfoni_2016} with a high spectral resolution. 
Therefore, we studied the stability of hydrohalite not only at relevant pressures and temperatures, as described in the first part of this study but also under irradiation with 2 keV electrons. 
We found that the depth of the band decreased by more than 50\% when irradiated with a dose corresponding to a few years on the surface of  Europa. This shows that the hydrohalite forming during the sublimation of the top layer is continuously dehydrated by the \SI{2}{\kilo\electronvolt} electrons. The dehydration happens on shorter timescales than the hydration of NaCl through sublimation, which is consistent with the non-detection of the band in observations. However, local thermal anomalies, such as active hotspot regions, could shift this equilibrium in favour of the hydration driven by sublimation. A surface temperature of \SI{145}{\kelvin} would be sufficient so that the hydration time scale is comparable to the irradiation-induced dehydration timescale. Such hotspots up to a size of \SI{100}{\kilo\meter^2} can not be ruled out by existing remote sensing data \citep{rathbun_galileo_2010}. The Europa Thermal Emission Imaging System (E-THEMIS) onboard Clipper \citep{christensen_europa_2024} will be able to detect such thermal hotspots if they are present on the surface of Europa. 
The color centers forming in the sample with 5wt\% NaCl are much stronger than the color centers observed within Tara Regio by \citet{trumbo_sodium_2019}, indicating that the abundance of NaCl on Europa's surface is lower than 5wt\%. 

\subsection{Implications for the interpretation of remote sensing data}
Sublimation changes the optical properties of a sublimating icy surface like the one of Europa on geologically short timescales. Calculated ablation rates are higher but in the same order of magnitude as sputtering rates estimated by \citet{cooper_energetic_2001}. Therefore, sublimation must be addressed when quantitative predictions about bulk composition from remote sensing data are being made. If the effect of sublimation is neglected, the interpretation of remote-sensing data is prone to overestimate the bulk abundance of salts in the surface ice layers on Europa.  This will be of special interest in interpreting the wealth of remote-sensing data expected in the coming decade from optical instruments like MISE and EIS on Europa Clipper as well as MAJIS and JANUS on JUICE. Our results indicate that hydrohalite is not stable on the surface of Europa, so the \SI{1.98}{\micro\meter} feature is not a useful measure of NaCl abundance. A local detection of hydrohalite on the surface of Europa would indicate a very fresh surface ($<10$ yr) or a thermal anomaly ($>145$ K), such as an active hotspot.

\section{Summary}
\label{sec:summary}
The Vis-NIR reflectance of particulate ice analogs containing three different salt species during sublimation has been studied. A novel method to measure the surface temperatures of the sample allowed the scaling of the sublimation kinetics using a simple thermal model based on measured surface brightness temperatures from Galileo data. Therefore, the time required to observe equivalent changes on Europa can be estimated. For all analog samples, we observe a rapid reduction of the broad water absorption bands around \SI{1.5}{\micro\meter} and \SI{2}{\micro\meter} over a time of a few thousand years at equatorial conditions on Europa. This impacts the geometric albedo in the order of 10\%, significantly changing the thermal balance of the surface. The Mg-bearing samples show skewed water absorption bands on the same timescale. During the sublimation, the analog sample produced from a NaCl solution developed a narrow absorption feature at \SI{1.98}{\micro\meter} and a shoulder at \SI{1.8}{\micro\meter} attributed to hydrohalite. We studied the stability of this feature under \SI{2}{\kilo\electronvolt} electron irritation and observed significant dehydration when irradiated with a dose equivalent to a few years on Europa. Therefore we expect that any hydrohalite present on the surface of Europa rapidly dehydrates, so a local detection would indicate a fresh surface and/or thermal anomaly, both signs of recent activity.

\section*{Data Availability}

The datasets supporting the findings of this study are available at the SSHADE database:

\begin{itemize}
  \item \url{https://doi.org/10.26302/SSHADE/EXPERIMENT_RO_20240312_001}
  \item \url{https://doi.org/10.26302/SSHADE/EXPERIMENT_RO_20240328_000}
  \item \url{https://doi.org/10.26302/SSHADE/EXPERIMENT_RO_20240405_000}
  \item \url{https://doi.org/10.26302/SSHADE/EXPERIMENT_RO_20240701_000}
\end{itemize}

\section*{Acknowledgements}
This work has been carried out within the framework of the National Centre of Competence in Research PlanetS supported by the Swiss National Science Foundation under grant 51NF40\_205606. The authors acknowledge the financial support of the SNSF. The authors also want to thank the technical staff at the Department of Space Research and Planetary Sciences, namely Mathias Brändli and Iljadin Manurung, for their support during the development of the experiment.

\section*{Author contributions}
\textbf{RO:} Conceptualization, Methodology, Software, Formal analysis, Investigation, Visualization, Writing - Original Draft \textbf{AP:} Conceptualization, Resources, Data Curation, Writing - Review \& Editing, Supervision, Funding acquisition \textbf{LLS:} Conceptualization, Methodology, Writing - Review \& Editing \textbf{LO:} Investigation \textbf{AG:} Investigation, Writing - Original Draft \textbf{AM:} Conceptualization, Resources \textbf{PW:} Resources, Writing - Review \& Editing, \textbf{NT:} Resources, Writing - Review \& Editing

\appendix
\section{Calibration of the thermopile sensor for surface temperature measurements}
\label{sec:thermopile-calibration}
A custom calibration procedure and calibration algorithm were developed to enable surface temperature measurements of ice in vacuum at cryogenic temperatures. The calibration reference sample was a solid ice slab with a rough surface to mimic the scattering and emission characteristics of porous ice in the far infrared (FIR) wavelength range. Smooth bare ice would have a significantly lower emissivity \citep{hori_modeling_2013}. The temperature was measured with a resistive temperature sensor (PT1000; 4-wire) frozen in the slab. The good thermal conductivity of the solid ice slab (\SI{4.3}{\watt/\meter\kelvin} at \SI{150}{\kelvin}, \citet{slack_thermal_1980}) allows the assumption that the surface temperature of the ice slab is equal to the measured temperature within the ice (upper limit $\Delta T < $\SI{1}{\kelvin}).

\begin{figure}
    \centering
    \includegraphics[width=\linewidth]{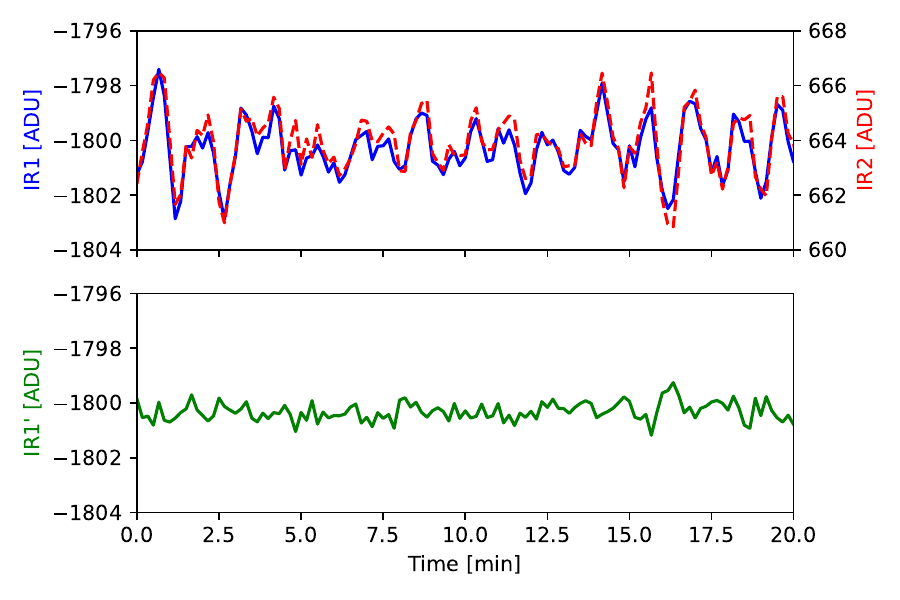}
    \caption{The upper panel shows that both thermopile elements recorded the same noise pattern caused by temperature fluctuations of the base. The deviation of the $IR_2$ channel with respect to its mean of the last 5min is calculated and removed from the $IR_1$ channel. The residual, "gradient corrected" signal can be seen in the lower plot.}
    \label{fig:ir-gradient-removal}
\end{figure}

The raw values from the thermopile analog/digital converters were read directly from the sensor using the I$^2$C protocol. This is necessary because the onboard calibration has a computation overflow at low temperatures, as we measure far outside of the specifications of the device. A custom calibration algorithm was developed to calibrate the surface temperature from the raw thermopile sensor output. The calibration involves 3 steps: 

The FIR sensor has two thermopile elements. One is in the focal point of the optics ($IR_1$), and the other one is placed off-axis ($IR_2$) for a so-called "gradient correction." Thermopile elements always measure a temperature difference between their surface and the sensor base. If the ambient temperature is lower than the sensor, the surface of the thermopile element cools down and a negative voltage is measured. Because of the thermal inertia of the elements, fluctuations in the temperature of the sensor base impact the measured voltage. Since both thermopile elements are affected, the off-axis element can be used to subtract a signal offset, as seen in Figure \ref{fig:ir-gradient-removal}. 
The gradient removed $IR_1'$ signal is calculated as
\begin{equation}
    IR_1' = IR_1 - (IR_2 - \overline{IR_2}^{5m}),
\end{equation}
where $\overline{IR_2}^{5m}$ is the simple moving average over the last 5min.
Note that this correction leaves averaged values ($>$ 5min window length) invariant.

The second step is the background signal subtraction. Long-term changes like the temperature of the shroud or vacuum chamber walls impact the measured signal of the thermopile.  We discovered that the off-axis thermopile element can be used as a proxy measurement for the ambient far-infrared background.

A simple toy model helps to put the amount of reflected background radiation on the sample into perspective: The intensity of thermal infrared radiation from an infinite sample with an emissivity $\epsilon=0.98$ at $T=\SI{110}{\kelvin}$ would be roughly equal to the intensity of the reflected ambient thermal radiation ($T_{amb}=\SI{293}{\kelvin}$) when no cooling shroud is present. In our chamber, this equality is shifted to lower temperatures because of the cooling shroud.

\begin{figure}
    \centering
    \includegraphics[width=\linewidth]{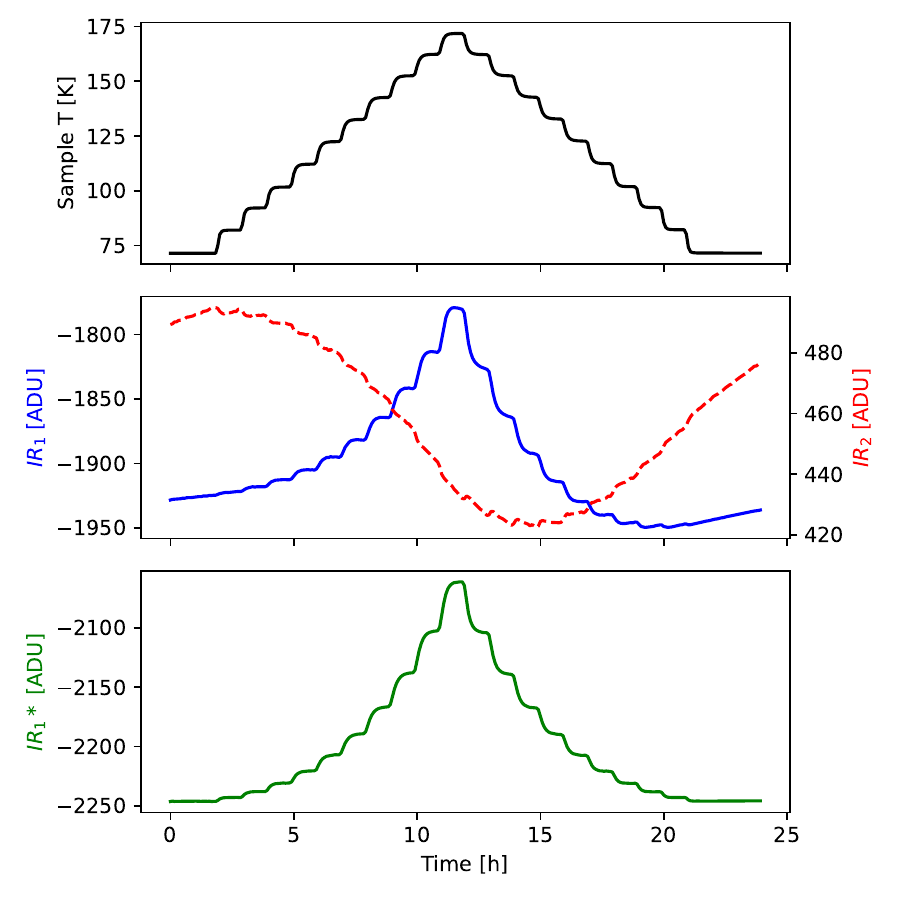}
    \caption{The top panel shows the temperature measured in the solid ice slab during the calibration profile. The temperature was changed in \SI{10}{\kelvin} increments, and between steps, equilibrium was reached. The second channel shows averaged outputs from IR channels 1 and 2. Changes in the FIR radiation background can be well measured with the second channel. In this run, the shroud temperature drifts by more than \SI{50}{\kelvin} because it is not directly cooled. Panel 3 shows the IR signal after the background subtraction using the second channel.}
    \label{fig:fir-background-profile}
\end{figure}

The corrected signal $IR^*_{1}$ is calculated as
\begin{equation}
    IR^*_{1} = IR_{1}' -  f_{bg} * \overline{IR_2}^{5m},
\end{equation}
where $f_{bg}$ is a background subtraction factor optimized to minimize offsets caused by changes in the ambient FIR background. In our chamber, the factor is 0.65 for a warm shroud and 0.6 for a cooled shroud. Figure \ref{fig:fir-background-profile} shows a calibration profile. The sample was heated in increments of \SI{10}{\kelvin}. 

In the last step, a third-order polynomial is fitted to the 10s-averaged preprocessed signal $IR^*_{1}$. A root-finding algorithm is used to retrieve the temperature as a function of $IR^*_{1}$ in the following measurements with porous ice.

\begin{figure}
    \centering
    \includegraphics[width=0.8\linewidth]{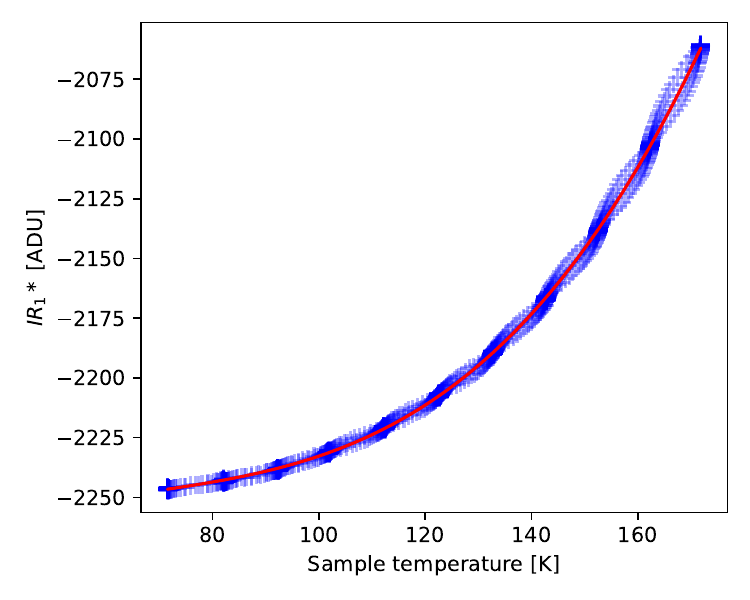}
    \caption{The blue markers show the  1min-averaged $IR_1^*$ signal during a calibration profile measurement. The red line shows a third-order polynomial fit of the data. This fit is used in other measurements to calibrate the surface temperature from the preprocessed $IR_1^*$ data. The bulges above and below the line show the phases when the sample temperature increases or decreases and the system is not in equilibrium. The dark blue patches show equilibrium states when the temperature is stable for a longer period.}
    \label{fig:ir-sample-temp}
\end{figure}

\begin{figure}
    \centering
    \includegraphics[width=0.8\linewidth]{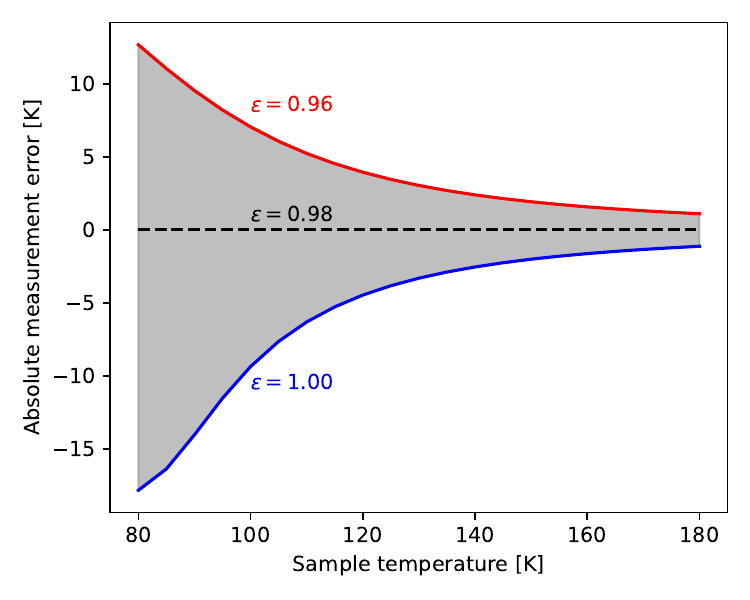}
    \caption{The absolute measurement error for a sample with a varying emissivity in function of temperature caused by the changing amount of reflected ambient radiation. The calculation is based on the measured signal offset when the sample is shielded from the ambient background radiation. We assume an emissivity $\epsilon=0.98$ for the calibration standard.}
    \label{fig:ir-error}
\end{figure}

The biggest source of uncertainty comes from the assumption that the emissivity of the calibration target is the same as the one of the measured sample. Therefore we estimate the measurement error caused by a varying emissivity. This can be done by measuring the offset of the measured FIR signal when the polished lid on the shroud is closed and the sample is shielded from the ambient FIR radiation. In the simulation chamber used in this study, the measured offset is $\Delta_{IR_1^*}=6$. From this offset, we can calculate the measurement error caused by the varying amount of reflected ambient radiation for other emissivities, assuming the emissivity of the calibration target was $\epsilon=0.98$. Figure \ref{fig:ir-error}
shows the absolute measurement error in the function of temperature for sample emissivities $\epsilon=0.96-1$. Pure ice samples with various surface morphologies have emissivities in this range \citep{hori_modeling_2013}. 

\bibliographystyle{elsarticle-harv} 
\bibliography{references}



\end{document}